\documentclass{aa}
\usepackage[utf8]{inputenc}
\usepackage{url}
\usepackage{tikz}
\usetikzlibrary{positioning, shapes.geometric}
\usepackage{multirow}
\usepackage{array}
\newcolumntype{P}[1]{>{\centering\arraybackslash}p{#1}}
\usepackage{footmisc}
\usepackage{color}
\usepackage[colorlinks = true,
           linkcolor = blue,
            urlcolor  = blue,
            citecolor = blue,
            anchorcolor = blue]{hyperref}

\newcommand{\forse}{{\textsc{ForSE}}}

\newcommand{\forsep}{{\textsc{ForSE\raisebox{+0.25ex}{+}}}}
\newcommand{\fsth}{{\textsc{ForSE\raisebox{+0.25ex}{+}S3}}}
\newcommand{\fstw}{{\textsc{ForSE\raisebox{+0.25ex}{+}S12}}}
\newcommand{\forses}{{\textsc{ForSE\raisebox{+0.25ex}{+}S}}}
\newcommand{\forsed}{{\textsc{ForSE\raisebox{+0.25ex}{+}D}}}
\newcommand{\planck}{{\it Planck}}

\usepackage{amsmath}
\defcitealias{2021ApJ...911...42K}{KP2021}

\begin{document}
\title{\forsep{}: Simulating non-Gaussian CMB foregrounds at 3 arcminutes in a stochastic way based on a generative adversarial network}

\author{Jian Yao \orcid{0000-0003-0813-9480} \inst{1,2,3} , Nicoletta Krachmalnicoff \orcid{0000-0002-5501-8449} \inst{1,2,3}, Marianna Foschi \orcid{0000-0001-8147-4993} \inst{4}, Giuseppe Puglisi \orcid{0000-0002-0689-4290} \inst{5,6,7} , Carlo Baccigalupi \inst{1,2,3}}

\institute{SISSA, Via Bonomea 265, 34136 Trieste, Italy  \\
\email{jyao@sissa.it}
\and INFN, Via Valerio 2, 34127 Trieste, Italy 
\and IFPU, Via Beirut 2, 34014 Trieste, Italy 
\and Instituto de Astrofísica de Andalucía, Glorieta de la Astronomía s/n, 18008 Granada, Spain 
\and Dipartimento di Fisica e Astronomia, Universit\`a degli Studi di Catania, via S. Sofia, 64, 95123, Catania, Italy 
\and INFN - Sezione di Catania, Via S. Sofia 64, 95123 Catania, Italy 
\and INAF - Osservatorio Astrofisico di Catania, via S. Sofia 78, 95123 Catania, Italy}
         
\abstract{
    We present \forsep{}, a Python package that produces non-Gaussian diffuse Galactic thermal dust emission maps at arcminute angular scales and that has the capacity to generate random realizations of small scales. This represents an extension of the \forse{} (Foreground Scale Extender) package, which was recently proposed to simulate non-Gaussian small scales of thermal dust emission using generative adversarial networks (GANs). With the input of the large-scale polarization maps from observations, \forsep{} has been trained to produce realistic polarized small scales at $3\arcmin$ following the statistical properties, mainly the non-Gaussianity, of observed intensity small scales, which are evaluated through Minkowski functionals. Furthermore, by adding different realizations of random components to the large-scale foregrounds, we show that \forsep{} is able to generate small scales in a stochastic way. In both cases, the output small scales have a similar level of non-Gaussianity compared with real observations and correct amplitude scaling as a power law. These realistic new maps will be useful, in the future, to understand the impact of non-Gaussian foregrounds on the measurements of the cosmic microwave background (CMB) signal, particularly on the lensing reconstruction, de-lensing, and the detection of cosmological gravitational waves in CMB polarization B-modes.
    }


\titlerunning{\forsep{} Simulating stochasitc non-Gaussian CMB foregrounds}
\authorrunning{Yao et al.}

\maketitle
% \tableofcontents
\nolinenumbers % comment for line numbers

\section{Introduction}

The cosmic microwave background (CMB) contains valuable information about the initial conditions in the early Universe and physical processes that happened during its propagation from the last scattering surface to our observation \citep{2002ARA&A..40..171H}. In addition to the great success of observations and interpretation of the CMB total intensity signal, mainly sourced  by scalar perturbations \citep{2013ApJS..208...19H, 2020A&A...641A...6P}, the focus of experimental efforts has now been transferred to the observations of the CMB polarization signal. In particular, the quest for the curl or B-mode component in the CMB polarized signal has drawn the most attention nowadays, as it can be a unique probe of primordial gravitational waves (PGWs, i.e., tensor-mode cosmological perturbations) \citep{1997PhRvD..55.1830Z, 1997PhRvD..55.7368K, 2016ARA&A..54..227K}. These tensor perturbations are predicted by the families of inflationary theories, which describe a history of accelerated expansion in the very early Universe \citep{1981PhRvD..23..347G}. Great efforts are being made with many ongoing and future telescopes, including ground-based ones, such as PolarBear\footnote{Polarization of the Background Radiation experiment} \citep{2022ApJ...931..101A}, SPT\footnote{South Pole Telescope} \citep{2023PhRvD.108b3510B}, ACT\footnote{Atacama Cosmology Telescope} \citep{2023arXiv230405203M}, SO\footnote{Simons Observatory}\citep{2019JCAP...02..056A}, CMB-S4\footnote{CMB-Stage-IV} \citep{2016arXiv161002743A}, BICEP\footnote{Background Imaging of Cosmic Extragalactic Polarization} \citep{2022ApJ...927...77A}, and AliCPT\footnote{Ali CMB Polarization Telescope} \citep{10.1093/nsr/nwy019}, balloon-borne ones such as SPIDER\footnote{Suborbital Polarimeter for Inflation Dust and the Epoch of Reionization} \citep{2022ApJ...927..174A}, and telescopes on satellites, like LiteBIRD\footnote{Lite (Light) satellite for the studies of B-mode polarization and Inflation from the cosmic background Radiation Detection} \citep{2023PTEP.2023d2F01L}. The tightest constraint on the amplitude of the PGW, in terms of the tensor-to-scalar ratio, $r \equiv A_{T}/A_{S}$,  is $r < 0.032$ \citep{2022PhRvD.105h3524T}, at a $95\%$ confidence level.

While the scalar perturbations do not contribute to the CMB B-modes and only source the parity-even E-modes, the  gravitational lensing effect from the forming cosmological structures along the propagation of CMB photons, with a typical deflection angle of a few arcmins, converts a small portion of E-modes into B-ones. These lensing-induced B-modes dominate over the PGWs at arcminute angular scales \citep{2000PhRvD..62d3007H, 2020A&A...641A...8P}. Some inflationary models also predict nearly Gaussian statistics of the CMB anisotropies with small deviations known as primordial non-Gaussianity (PNG), which can be used to further constrain the vast space of inflationary theories that go beyond a simple single field slow roll scenario \citep{2004PhR...402..103B, 2020A&A...641A...9P}. However, the lensing effect will also induce non-Gaussianity into the distribution of CMB photons by distorting the paths and coupling different multipole bins of power spectra. 

Despite being a contaminant to PGWs and PNG, CMB lensing carries cosmological information on processes occurring during structure formation, on neutrino masses \citep{2015PhRvD..92l3535A}, and Dark Energy \citep{2006PhRvD..74j3510A}. Effective techniques capable of reconstructing the lensing field have been developed, mainly relying on detecting the non-Gaussianity presented in the observed CMB maps \citep{2002ApJ...574..566H, 2021PhRvD.103h3524M}. The latest detection of lensing potential comes from the ACT experiment, which achieves a $43\sigma$ detection of the CMB lensing signal \citep{2023arXiv230405202Q}. 

The primary source of contamination to CMB B-modes arises from astrophysical processes causing diffuse emissions within our own Galaxy. The two main polarized components are represented by the thermal dust and synchrotron emission, dominating at high ($\gtrsim70$ GHz) and low ($\lesssim70$ GHz) frequencies, respectively, exceeding the CMB B-mode signal in any frequency and any position on the sky \citep{2016A&A...588A..65K, 2018A&A...618A.166K}. They are caused by dust grains heated by starlight, exhibiting a quasi-thermal emission at about 18 K, and cosmic ray electrons spiraling around the lines of the Galactic magnetic field, respectively. The class of data analysis algorithms capable of extracting the cosmological B-mode anisotropies out of a multi-frequency dataset is known as component separation. Several implementations exist, exploiting the different properties that CMB and foregrounds have, such as frequency dependency and spatial distribution. Depending on the assumptions made about the foregrounds, these methods can generally be classified into blind \citep{2018ApJS..239...36Y, 2021A&A...650A..65S, 2019MNRAS.484.1616Z, 2009A&A...493..835D} and parametric ones \citep{2009MNRAS.392..216S, 2020A&A...641A...4P, 2023A&A...675A...1B}.

Unlike the CMB, diffuse foregrounds generically exhibit a large degree of non-Gaussianity. This comes primarily from large-scale observations \citep{2019JCAP...10..056C, 2019A&A...629A.115A} and is expected to also be true at smaller scales, since dust grains are highly locally distributed and the magnetic field in the diffuse interstellar medium is highly turbulent. The foreground non-Gaussianity introduces coupling between different modes in the angular power spectra, which induces non-diagonal terms in the covariance matrices of the observed signal. On the other hand, covariance matrices are usually treated as diagonal under the assumption that all the sky components behave like Gaussian random fields. Additionally, approaches relying on power spectra will inherently ignore the non-Gaussianity, as the two-point functions are insufficient in capturing its presence. Therefore, the existence of non-Gaussianity in the Galactic foregrounds will potentially induce systematic errors in primordial B-mode detection. 

\cite{2023arXiv230909978A} claim that the constraints on $r$ are robust against the non-Gaussianity in the polarized dust at the power spectra level, but they only focus on $30 \leq \ell \leq 300$, the angular scales at which the PGWs cause a maximum B-mode contribution known as recombination bump. At smaller scales where lensing dominates, non-Gaussian foreground will possibly cause bias to lensing reconstruction \citep{2020JCAP...06..030B} and de-lensing. Moreover, non-Gaussianity also affects PNG measurements \citep{2010MNRAS.405..961C}.

The main reason for neglecting non-Gaussianity is that observations of foregrounds at arcminute scales are lacking, and so is our capability of generating realizations of simulations of their patterns. Current observations of polarized foregrounds are limited to degree scales over a large sky fraction and can only reach a higher resolution at sub-degree or arcminute scales for portions of the sky \citep{2020A&A...641A..11P, 2004MNRAS.351..436B, 2015MNRAS.451.4311R}.

The way to simulate foregrounds relies on data and phenomenology. Foreground observations at a given frequency are used as templates and extrapolated to other frequency bands by making assumptions about their spectral energy distributions (SEDs). These templates are mostly based on the observations by the \planck{} \citep{2020A&A...641A...4P} and Wilkinson Microwave Anisotropy Probe (WMAP) satellite observations \citep{2015MNRAS.451.4311R, 2013ApJS..208...20B}, which characterize the signal down to a limited angular resolution (around $1^{\circ}$). There exist several widely used packages that simulate Galactic foreground maps, such as the Python Sky Model ($\texttt{PySM}$, \cite{2017MNRAS.469.2821T}) and \planck{} Sky Model \citep{ 2013A&A...553A..96D}. The latest version of $\texttt{PySM3}$ package \citep{Zonca_2021, panex} uses foreground templates from the latest available observations and extrapolates them to arcminute angular scales. These packages usually fit a power law to the observed large-scale power spectra of foregrounds and extrapolate to small scales with Gaussian realizations of them. Although commonly used, these packages are not able to produce foreground small scales with significant non-Gaussianity comparable to the observed large scales. 

\cite{2019ApJ...887..136C} provides a data-driven framework to construct three-dimensional Stokes parameter maps of thermal dust emission in position–position–velocity space using only neutral hydrogen data on the basis that HI is strongly correlated with the dust in the diffuse interstellar medium \citep{2017ApJ...846...38L}. By integrating over the velocity space, they obtained the polarized dust emission maps over the full sky at a resolution of $16.2\arcmin$, corresponding to that of HI data, which are in good agreement with the \planck{} observed 353 GHz dust maps in terms of several physical properties such as the polarization fraction and power spectra, although they do not have a thorough discussion about the non-Gaussianity contained in their maps. 

\cite{2022ApJ...928...65H} uses a large number of filaments in the distribution of the thermal dust grains together with the large-scale template from data to reproduce the main features of the \planck{} 353 GHz map. These include the power spectrum slopes of intensity and polarization maps, the ratios between EE, BB, and TE power spectra, and the level of non-Gaussianity in the total intensity map, which can be controlled by the density of filaments. When focusing on scales of arcminutes or tens of arcminutes, $\ell = 300-1200$, however, the \planck{} 353 GHz total intensity map has more non-Gaussianity with respect to the generated small scales from the model based on filaments.

Diffuse foreground models are also generated by exploiting magnetohydrodynamic (MHD) simulations. They model the physical processes of the interstellar medium, such as heating and cooling of the gas and the interaction between dust grains and the turbulent magnetic field. The thermal dust emission can be obtained by integrating the emission of dust grains along the line of sight \citep{2001ApJ...559.1005P, 2015A&A...576A.105P, 2019ApJ...880..106K}. The simulated maps are thus non-Gaussian because of the MHD processes and can reach small scales according to the resolution of the MHD simulation itself. However, the MHD simulations can only reproduce the statistical properties of the Galaxy, and thus fail to generate the specific morphology of Galactic foreground emission. Another important limiting factor of MHD simulations is that they are computationally expensive, especially in achieving a high resolution.

An additional technique consists of simulating the foreground by exploiting innovative algorithms capable of modeling the emission pattern at a high resolution on the basis of the observed one at moderated and low angular scales. The \forse{} model introduced by \cite{2021ApJ...911...42K} (hereafter \citetalias{2021ApJ...911...42K}) is able to generate small scales, up to $12\arcmin$, starting from the low-resolution polarized dust observations. It utilizes a generative adversarial network (GAN), which is trained to inject small-scale features with statistical properties like the ones observed at a high resolution in the intensity maps.

In this study, we further extend the \forse{} algorithm  to: i) produce dust polarization maps at $3\arcmin$ resolution, which is vital for checking the stability of lensing reconstruction and de-lensing algorithms in CMB observations, ii) stochastically generate multiple realizations of small-scale features, which are most important in estimating the uncertainty associated with foreground variations.

Additional uses of deep generative models for dust simulations exist in the literature. For example, \cite{2021MNRAS.500.3889A} use GANs to generate simulated total intensity maps from the observed \planck{} generalized needlet internal linear combination (GNILC) map at 353GHz. Other common generative models, such as variational autoencoders \citep{2021MNRAS.504.2603T} or diffusion models \citep{2023arXiv231016285H}, are also used.

The outline of the paper is as follows. In Section \ref{sec:ForSE+}, we summarize the first version of the \forse{} algorithm and present the extension and methodology used to achieve the new version operating at arcminute scale. In Section \ref{sec:pre-post}, we describe the pre-processing, training, and post-processing of the new model and also present our validation procedure. In Section \ref{sec:full-sky}, full-sky maps are presented and compared with maps from the latest version of $\texttt{PySM3}$ package. Conclusions are given in Section \ref{sec:conclusions}.

\section{The \forsep{} model}
\label{sec:ForSE+}
In this section, we briefly review the basic structure and assumptions of the \forse{} algorithm presented in \citetalias{2021ApJ...911...42K}. We then introduce our new version of the code, \forsep{}, which allows one to generate maps of the thermal dust emission with non-Gaussian structures at $3\arcmin$ angular resolution, in both a deterministic and a stochastic way.

\subsection{Review of the \forse{} model: From $80\arcmin$ to $12\arcmin$} 
\label{sec:ForSEold}

As has already been mentioned, the \forse{} model, introduced and validated in \citetalias{2021ApJ...911...42K}, is based on GANs and allows one to produce non-Gaussian full-sky maps of polarized thermal dust emission at an angular resolution of $12\arcmin$ from low-resolution \planck{} observations at $80\arcmin$. 

Generative adversarial networks are a particular family of networks \citep{2014arXiv1406.2661G} whose characteristic feature is to be composed of two sub-networks called Generator ($G$) and Discriminator ($D$), which are trained to compete against each other. In practice, the goal of $G$ is to produce new images that are compared by $D$ with a set of real images that are the training set. Once the training of the two sub-networks is done in an adversarial way, $G$ is able to produce images that have the same statistical properties as those belonging to the training set, in such a way that mock and real images are no longer distinguishable by $D$.

In the \forse{} implementation, the input to $G$ are images at a low resolution ($80\arcmin$) of the thermal dust emission observed by the \planck{} satellite and processed through the GNILC method\footnote{\url{http://pla.esac.esa.int/pla/aio/product-action?MAP.MAP_ID=COM_CompMap_IQU-thermaldust-gnilc-unires_2048_R3.00.fits}} \citep{2020A&A...641A...4P}. The output of $G$ are small-scale features at $12\arcmin$, which are then compared by $D$ with real observations in total intensity at the same angular resolution. We note that different training is performed for total intensity and Stokes $Q$ and $U$ maps, but always having as the training set images of small scales $\tilde{m}_{12'}^{I, 20^{\circ}}$ in total intensity. \citetalias{2021ApJ...911...42K} therefore assumed that thermal dust statistical properties of small scales in polarization are the same as for Stokes $I$ maps. We also rely on that assumption in this work.

In \citetalias{2021ApJ...911...42K} and in this work, the following definition of “small scales” is used. Let $M_{TOT}$ be a foreground map at a given angular resolution, which can be seen as the sum of two maps containing, respectively, only large- and small-scale structures:
\begin{equation}
    M_{TOT} = M_{LS} + M_{SS} .
\end{equation}

Assuming that the map encoding small-scale features, $M_{SS}$, is modulated by the large scales, $M_{LS}$ -- that is, $M_{SS} = M_{LS}\cdot m_{ss}$ -- we have 
\begin{equation}\label{eqn:ss}
    M_{TOT} = M_{LS} + M_{LS}\cdot m_{ss} = M_{LS}\cdot \tilde{m}_{ss},
\end{equation}
where $\tilde{m}_{ss} = m_{ss} + 1$ represents the small-scale map generated by network $G$, having as an input $M_{LS}$.

Although there exist neural networks designed to work on the sphere \citep{2019A&A...628A.129K}, our GAN deals with flat two-dimensional images. For this reason, we had to project the input maps onto flat patches, and project output patches back onto the sphere, after the application of the trained GAN model. We used the same projection strategy as that described in the appendix of \citetalias{2021ApJ...911...42K}, which divides the GNILC thermal dust Stokes $Q$ and $U$ maps at 80$\arcmin$ with $N_{side} = 2048$\footnote{Presented in the \texttt{Hierarchical Equal Area Latitude Pixels (HEALPix)} \citep{2005ApJ...622..759G} format.} into 174 square patches that have $320\times320$ pixels and a physical side length of $20^{\circ}$. We note that this projection on flat patches and reprojection on the sphere can introduce distortions in the final full-sky map. In order to mitigate this effect, the flat patches overlap each other. We estimate that the final level of distortion induced by our procedure is always less than 7\% of the signal (around 2\% on average).

From now on we will use $\tilde{m}_{z'}^{X, y^\circ}$ and $M^{X, y^\circ}_{z'}$ to refer to the small-scale (output of the network) and large-scale (input of the network) patches, respectively, where $X=I/Q/U$ defines the Stokes parameters, $y^{\circ}$ the physical dimension of the patch in degrees, and $z\arcmin$ its angular resolution in arcminutes.

The GAN architecture, training procedure, input maps, and output validation of the \forse{} model are fully described in \citetalias{2021ApJ...911...42K}. 

\begin{figure*}[hbt]
    \centering
    \includegraphics[width=\textwidth, trim={0.075cm 0.1cm 0.075cm 0.1cm},clip]{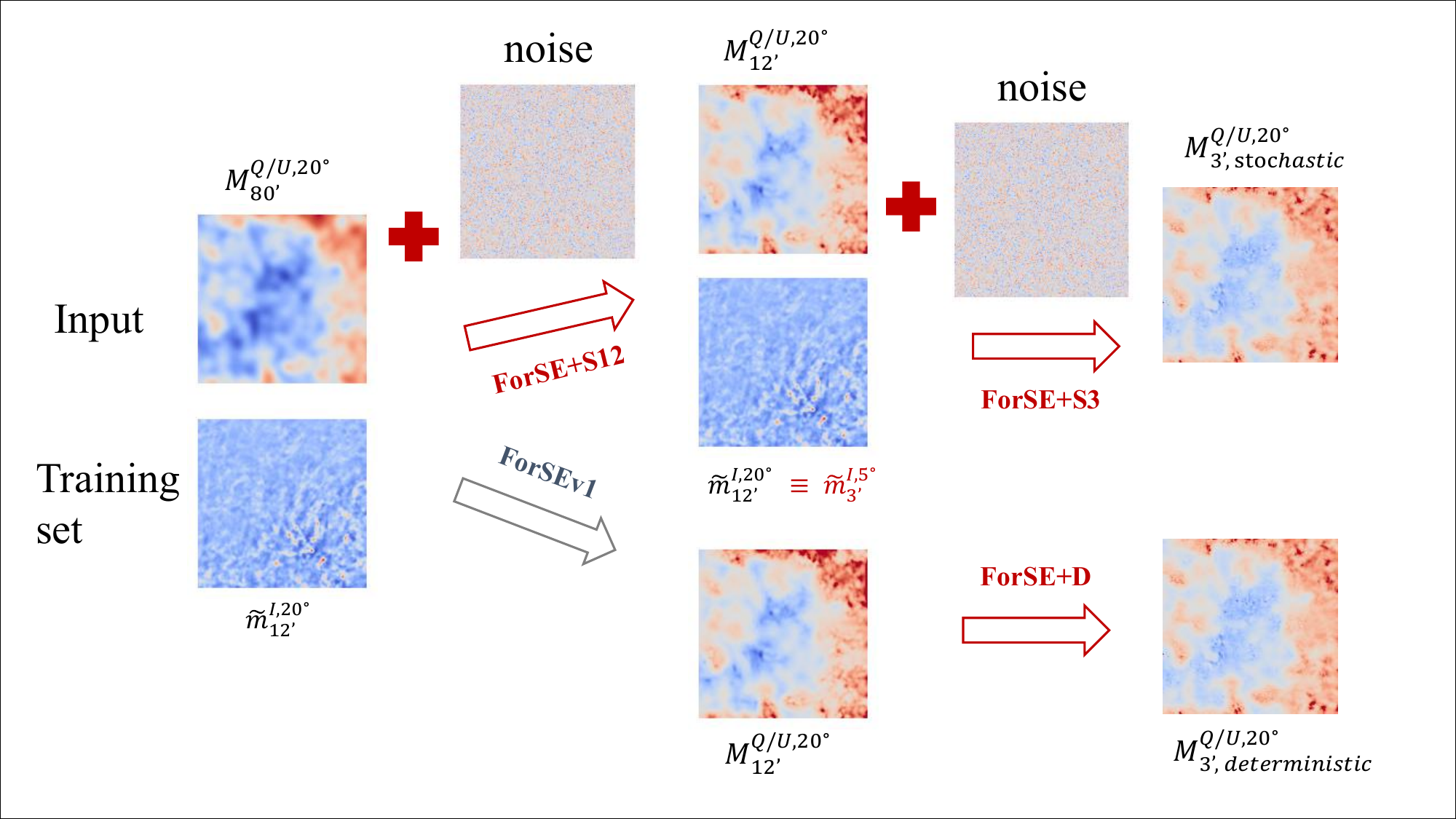}
    \caption{Structures in \forse{} (\citetalias{2021ApJ...911...42K}) and \forsep{} (this work). \forse{} is designed to achieve a
    $12\arcmin$ resolution from input large scales at $80\arcmin$. We have three kinds of newly trained models here: \fstw{} and \fsth{} to produce stochastic maps at $12\arcmin$ and $3\arcmin$,  and \forsed{} to generate a deterministic map at $3\arcmin$. The output, $M_{12\arcmin}^{Q/U, 20^\circ}$, from \forse{} and \fstw{} are the input to \forsed{} and \fsth{} to get deterministic and stochastic small scales at $3\arcmin$, respectively.}
    \label{fig:forse_forse_plus}
\end{figure*}

\subsection{\forsep{}: Producing non-Gaussian dust maps at $3\arcmin$}
\label{sec:ForSEnew}

In this work, we implement \forsep{}, an updated version of the \forse{} code and its training procedure, with two objectives:
\begin{enumerate}
\item To allow the generation of full-sky polarization maps with non-Gaussian features at an enhanced resolution of $3\arcmin$.
\item Starting from the same low-resolution maps, to generate multiple realizations of stochastic small scales that still have the correct non-Gaussian statistical properties. 
\end{enumerate}
In the sections below, we explain the assumptions and methodology used to achieve these two goals. Figure \ref{fig:forse_forse_plus} sketches the input and output of the new \forsep{} models and in Table \ref{tab:summary} we summarize the new models trained in this paper. 

\begin{table*}
    \centering
    \begin{tabular}{cP{0.4\linewidth}P{0.4\linewidth}}
    % \toprule
        Model & Description (Output) & Input\\
        \hline
        \multirow{2}{*}{\forse{}} & \citetalias{2021ApJ...911...42K} version to simulate deterministic thermal dust emission maps at $12\arcmin$ & \multirow{2}{*}{\planck{} GNILC maps at $80\arcmin$}\\
        \multirow{2}{*}{\fstw{}}&  new model to simulate stochastic thermal dust emission at $12\arcmin$ & \multirow{2}{*}{\planck{} GNILC maps at $80\arcmin$ plus random component}\\
        \multirow{2}{*}{\forsed{}}&  new model to simulate deterministic thermal dust emission maps at $3\arcmin$ & Maps at $20\arcmin$, smoothed from \forse{} $12\arcmin$ maps\\
        \multirow{2}{*}{\fsth{}}& new model to simulate stochastic thermal dust emission maps at $3\arcmin$ & Maps at $20\arcmin$ smoothed from \fstw{} plus random component\\
        \hline
    \end{tabular}
    \caption{Summary of three newly trained models in this paper and the first version of the model proposed in \citetalias{2021ApJ...911...42K}.}
    \label{tab:summary}
\end{table*}

In order to incorporate these extensions, we used the same GAN architecture as for the \forse{} model, re-implemented using the new 
Tensorflow\footnote{\url{https://www.tensorflow.org}} framework (version 2.6.0), and we performed additional fine-tuning steps of some hyper-parameters of the networks to improve our results. 

\subsubsection{Scale invariance assumption} \label{sec:scale-invariant}

The first goal of our implementation of \forsep{} is to generate a map of polarized dust emission with non-Gaussian features at arcminute angular scales. As we anticipated, this is crucial to understand the possible impact of non-Gaussianity in the extraction of the CMB $B$-mode lensing signals, which peak at these scales. In order to achieve this, we need to find a way to overcome the current limitation in the observational data at our disposal. As a matter of fact, in order to train our GANs we can only rely on a training set composed of 350 patches, with dimensions of $20^{\circ}\times20^{\circ}$ and $320\times320$ pixels, at an angular resolution of $12\arcmin$, taken from the total intensity GNILC \planck{} map at 353 GHz, described in \citetalias{2021ApJ...911...42K}. No other observations of thermal dust emission at a higher angular resolution in a portion of the sky large enough to be used as training set are available.

The idea to circumvent this restriction and still be able to reach a resolution higher than $12\arcmin$, even in the absence of a proper training set, is to make a scale-invariance assumption, applying our GAN model in an iterative way.
In practice, our set of training squared patches with dimensions of $20^{\circ}\times20^{\circ}$ and a resolution of $12\arcmin$ can equally be treated as having dimensions of $5^{\circ}\times5^{\circ}$ and resolution of $3\arcmin$, since the network does not have a sense of physical units and is only sensitive to the ratio between the dimension of the patch and its angular resolution (i.e., $\tilde{m}^{I, 20^{\circ}}_{12'}\equiv\tilde{m}^{I, 5^{\circ}}_{3'}$). In this way, the same dataset used to train the first version of \forse{} in \citetalias{2021ApJ...911...42K} can be used to train a new GAN model. This model takes the output of the first iteration of the code as input and generates non-Gaussian scales at $3\arcmin$. As has been mentioned, this implies that we are assuming scale invariance for the statistical properties of dust emission; that is, scales at $12\arcmin$ have the same properties as those at $3\arcmin$. This assumption is justified by the fact that current observations of the dust polarization power spectra shows that they can, at the first order, be approximated as a power law as a function of the angular scales \citep{2020A&A...641A..11P}. 

Additional pre- and post-processing of the input patches (including upsampling, smoothing, and the sub-division of patches) is needed to train the GAN in the correct way and to be able to restore full-sky maps, as is described in Section \ref{sec:pre-deter}.

\subsubsection{Stochasticity} \label{sec:stochasticity}

Our second goal is to produce different realizations of the non-Gaussian small-scale structures. This is important to estimate the variance of the signal we are simulating as well as the correlation among different angular scales. The way we achieved this was by simply adding a random component to the large-scale maps that are the input of our GANs. We then trained new models on these \emph{signal} \rm{+} \emph{random component} maps, always using the 350 $\tilde{m}_{12'}^{I, 20^\circ}$ patches as the training set.

The random component that we considered was simply generated as a random realization from a uniform distribution in the range $[-1,1]$. Since our input maps were always re-scaled to have pixel values ranging in the same interval to be compatible with the input of our neural networks, we had a signal-to-noise ratio ($\rm SNR)$ of $\sim1$. We refer to this as “stochastic training;” in other words, a random component was added to the input signal, \forses, as opposed to the deterministic case (\forsed{}), in which we did not add any random component to the input maps.

\section{Pre-processing, training, and post-processing of \forsep{}} \label{sec:pre-post}

In this section, we describe the training details of \forses{} and \forsed{}, including the pre-processing and post-processing steps, and present the results on flat-squared patches, including maps and power spectra, before reprojecting them to full-sky maps. 

\subsection{\forses{} to $12\arcmin$}

We first describe how we generate stochastic maps with non-Gaussian features at a resolution of $12\arcmin$. We also applied a similar procedure to construct maps at $3\arcmin$, as is described in Section \ref{sec:train_3}. For clarity, we will call \forses{}12 the model that generates stochastic maps at $12\arcmin$ and \forses{}3 the one that goes up to $3\arcmin$.

As was mentioned above, we injected stochasticity into our generative process by simply adding a random component to the low-resolution maps that are the input of our GAN. By doing so, \forses{} is able to generate non-Gaussian small-scale features that still partially depend on the real observed large-scale structures but will have a different morphology as we change the realization of the random component in the input dataset.

\subsubsection{Training and post-processing}

As in \citetalias{2021ApJ...911...42K}, we trained two models for $Q$ and $U$ maps separately. The inputs to the generator, $G$, were the 174 $M^{Q/U, 20^{\circ}}_{80\arcmin}$ patches that together cover the full sky, with dust signal plus an additional random component. The training set, which encodes the target statistical distribution of small scales, was the 350 $\tilde{m}^{I, 20^{\circ}}_{12\arcmin}$ maps. The weights of the neural works were updated for $2\times10^{5}$ epochs and saved every 500 steps. Since we do not want to generalize the usage of the trained model -- the trained model only needs to predict the output from the training data -- it is not a problem if there is an over-fitting during training. Therefore, there was no consideration of a separate validation set during the training process.

Following \citetalias{2021ApJ...911...42K}, we used the Minkowski functionals (MFs), as implemented in \citep{2008JSMTE..12..015M}, as the criteria with which to choose the best epoch. For two-dimensional fields, three kinds of MFs can be built: $\mathcal{V}_0$, $\mathcal{V}_1$, and $\mathcal{V}_2$. They fully describe the statistical properties of the field and represent the covered area ($\mathcal{V}_0$), the boundary length ($\mathcal{V}_1$), and the number difference of connected domains and holes of the image's feature ($\mathcal{V}_2$), as a function of the threshold, $\rho$ \citep{hadwigerVorlesungenUeberInhalt1957}.

During the GAN training process, the goal is to produce small-scale feature maps with statistical properties as close as possible to the ones of the training set.  We quantified the level of agreement by calculating the overlapping fractions between the distributions of MFs of the maps in the training set ($\tilde{m}^{I, 20^{\circ}}_{12\arcmin}$) and those of the generated ones ($\tilde{m}^{Q/U, 20^{\circ}}_{12\arcmin}$). The distributions of MFs are indicated by the $\pm 1 \sigma$ variation among the training set or generated maps and the overlapping fractions were computed as the ratio between the intersection area and the total area spanned by the two distributions. In practice, we calculated the MFs overlapping for each saved epoch of $G$ and selected as the best model the one with the highest score. In doing so, we computed the MFs for the output maps, $\tilde{m}^{Q/U, 20^{\circ}}_{12\arcmin}$, by applying $G$ to maps with the realization of a random component different from the one used for training.

\begin{figure*}[hbt]
    \centering
    \includegraphics[width=180mm]{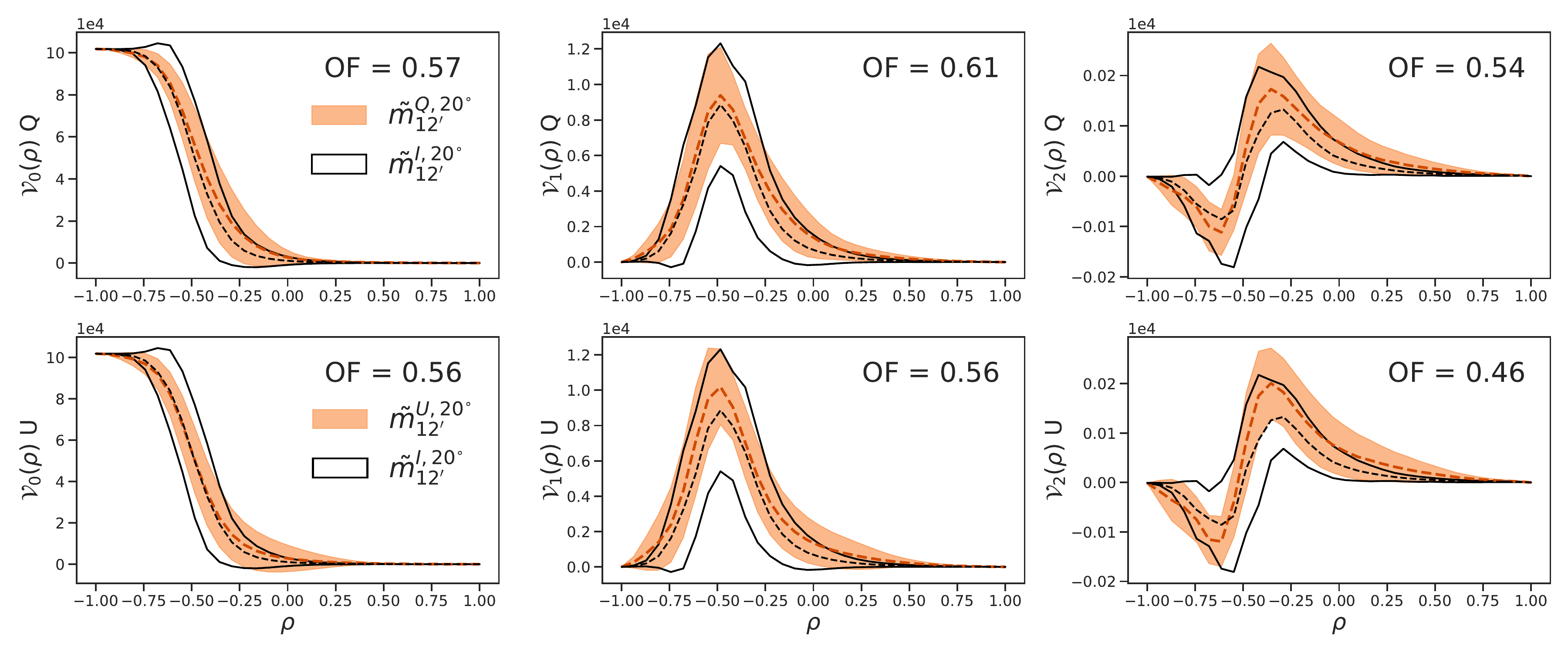}
    \caption{MFs of small scales at $12\arcmin$ produced by \forses{}, compared with the ones from the intensity maps in the training set. The overlapping fractions (OFs) of each pair of MFs are also shown. The dashed line represents the mean over the set of different patches and the shaded area is their standard deviation. The distribution of $Q$ maps is shown in the upper panel, and $U$ in the lower panel.}
    \label{fig:MFs_random_12}
\end{figure*}

The best models are obtained after 5500 epochs and 6000 epochs for $Q$ and $U$, respectively. Their MFs are shown in Figure \ref{fig:MFs_random_12} and compared with the ones from the training set. The overlap among the distributions is at a level of $50\%-60\%$, comparable with the one obtained in \cite{Krachmalnicoff_2023} (that includes corrections to \citetalias{2021ApJ...911...42K}). In comparison, Gaussian small scales have MFs with obviously different shapes from these two sets of maps, as is illustrated in Fig.7 of \citetalias{2021ApJ...911...42K}.

We also computed the overlapping fraction of the MFs by considering 100 different realizations of the small-scale maps (obtained by changing the random component in the input). The mean (standard variation) values of the overlapping fraction for $\mathcal{V}_0$, $\mathcal{V}_1$, and $\mathcal{V}_2$ are 59.1\%(1.5\%), 62.9\%(0.6\%), and 55.8\%(0.7\%), respectively, for Stokes $Q$ maps, and 58.8\%(2\%), 56.3\% (1.3\%), and 45.5\%(1.1\%), respectively, for $U$ maps. These numbers show the robustness of \forses{} in generating stochastic small scales with non-Gaussian high-order statistics. 

Since in the training procedure both the input maps and the training sets are normalized in the range of [-1,1], the output maps also have pixel values in this range, and therefore need to be normalized to restore physical units. We achieved this by following the procedure of \citetalias{2021ApJ...911...42K}, hence ensuring that, for each patch and for both $Q$ and $U$, the amplitude of the power spectrum of the produced small scales matches the extrapolation at higher multipoles of the power spectrum of the low-resolution input maps at $80\arcmin$. 

\subsubsection{Results of \forses{} at $12\arcmin$}

\begin{figure*}[hbt]
    \centering
    \includegraphics[width=0.85\textwidth]{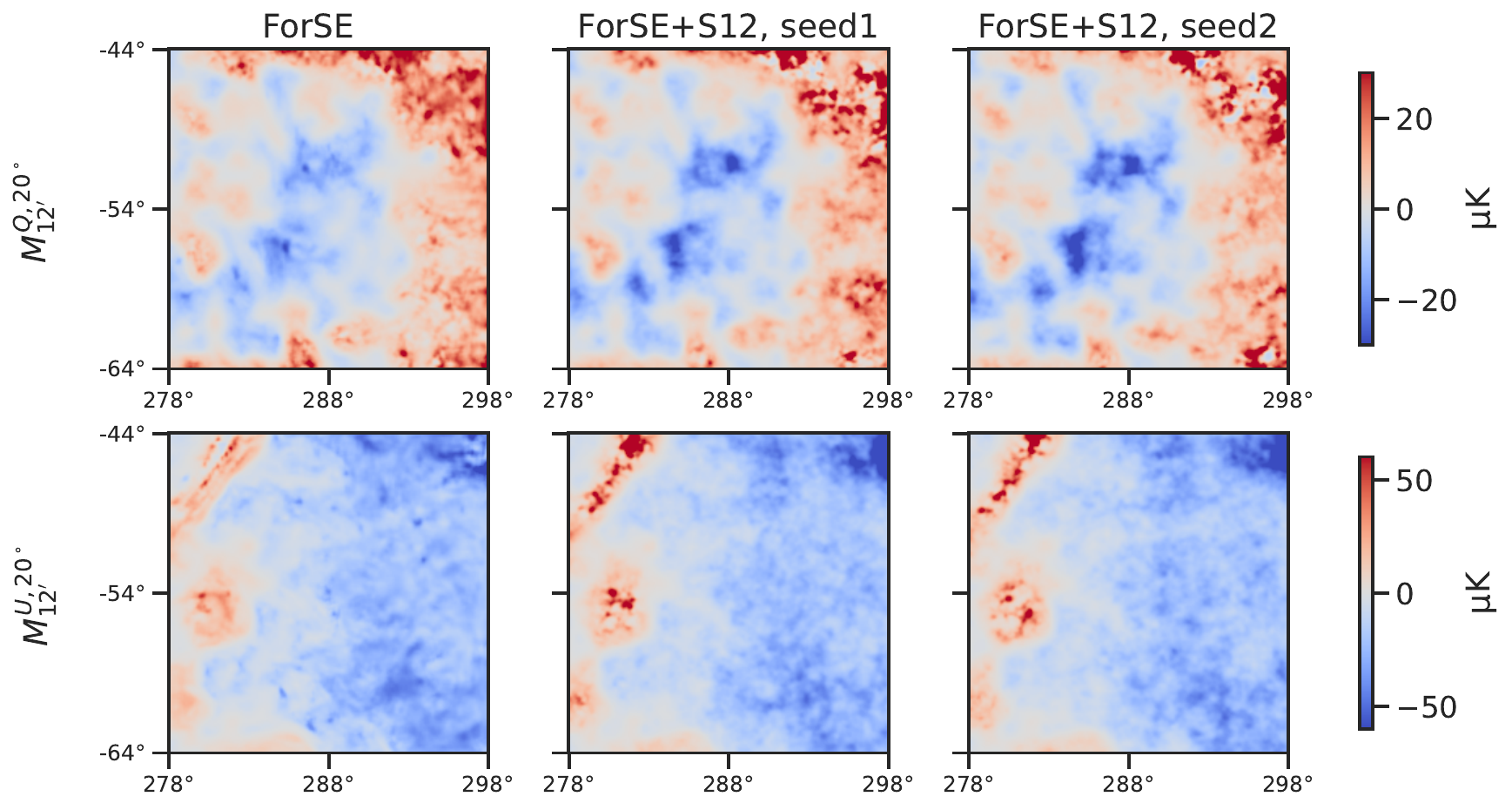}
    \caption{Maps of $20^\circ \times 20^\circ$ patches at $12\arcmin$. From left to right are the deterministic map from \forse{} and two stochastic realizations from \fstw{}. Throughout this paper, all the maps are shown in Galactic coordinates.}
    \label{fig:maps_12}
\end{figure*}

Figure \ref{fig:maps_12} shows $M_{12\arcmin}^{Q/U, 20^\circ}$ patches with two different realizations of small scales at $12\arcmin$ from \fstw{}, after the normalization mentioned above, in the second and third columns, compared with the deterministic results from \forse{} in the first column. Both the differences at small scales and consistency at large scales between the outputs from \forse{} and \fstw{} show the capability of the trained model to produce small-scale features in the map space.

\begin{figure*}[hbt]
    \centering
    \includegraphics[width=0.9\textwidth]{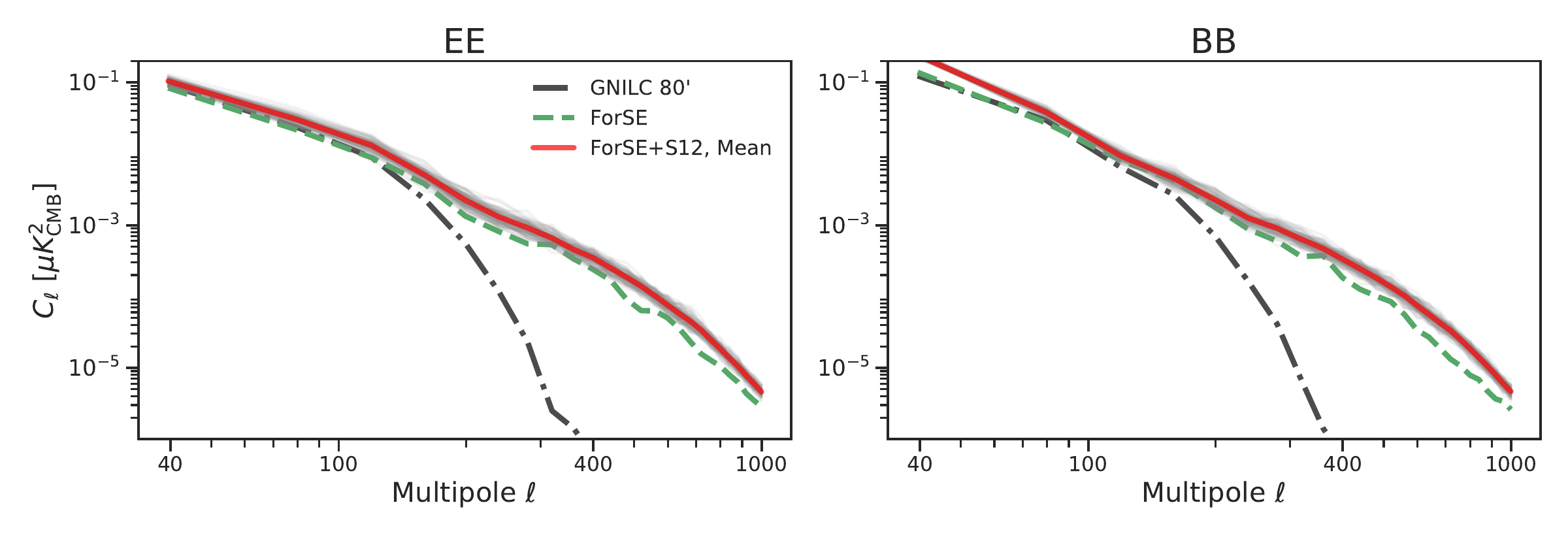} \caption{EE and BB power spectra of the squared patches at $12\arcmin$ shown in Figure \ref{fig:maps_12}. Dash-dotted black lines are the power spectra for GNILC $80\arcmin$ patches and green lines are for the deterministic $12\arcmin$ patches. Gray lines show the power spectra of all the 100 realizations from \fstw{} and red lines are the means. }
    \label{fig:ps_12}
\end{figure*}

To further validate these maps, we calculated the second-order statistics -- the power spectrum -- using the Namaster package \citep{2019MNRAS.484.4127A}.\footnote{\url{https://namaster.readthedocs.io/en/latest/sample_flat.html}} In Figure \ref{fig:ps_12}, we show the EE and BB power spectra from the $QU$ squared patches of low-resolution maps, the output from \forse{}, and 100 realizations of \fstw{} in black, green, and gray, respectively. The mean values of power spectra from 100 stochastic realizations are also shown in red. The output maps from \fstw{} are consistent with the deterministic output maps in terms of the power spectra, as was expected.

\subsection{\forsed{} and \fsth{} to $3\arcmin$} \label{sec:train_3}

\begin{figure*}[hbt]
    \centering
    \includegraphics[width=\textwidth, trim={0.075cm 0.1cm 0.075cm 0.1cm},clip]{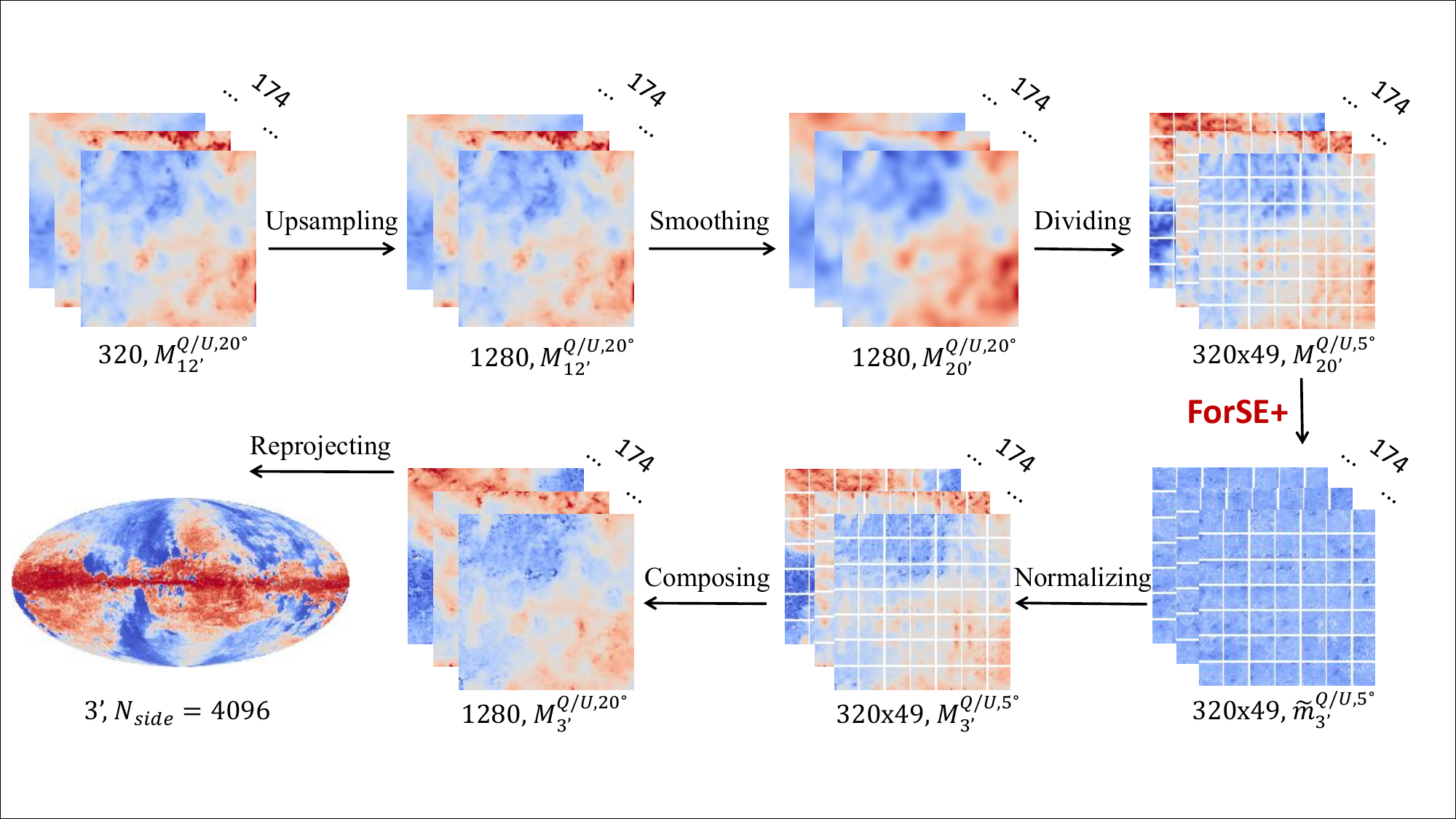}
    \caption{Pipeline to prepare the training data for \forsed{} (upper row, from left to right) and post-processing to transform the output patches from \forsed{} into the full-sky map at $3\arcmin$ in the \texttt{HEALPix} format (lower row, from right to left). 174 in the upper right corner of each tile means the number of images. The first number beneath the images indicates the number of pixels on each side of the image and the second expression has the same subscript and superscript as that described in Section \ref{sec:ForSEold}. We note that after the “Dividing” and before the “Composing” steps, each original image was divided into 49 sub-patches, with a side length of $5^\circ$. At the end, all the flat patches were reprojected onto a \texttt{HEALPix} map with $N_{side}$ equal to 4096.} 
    \label{fig:forse_flowchart}
\end{figure*}

We describe now the procedures that we followed to generate maps at the resolution of $3\arcmin$, by using our GAN model iteratively both in the case of \forsed{} and for \fsth{}. The whole procedures, including several pre- and post-processing steps, are sketched in Figure \ref{fig:forse_flowchart} and described in the following (see also \cite{marianna}).

\subsubsection{Pre-processing for the training of \forsed{}} \label{sec:pre-deter}

As was mentioned above, we reached the resolution of $3\arcmin$ by assuming scale invariance in the statistics of the thermal dust emission.  We exploited the output of the first GAN model as the input to a second generative step by using the same training set and considering it to be composed of 350 patches with dimensions of $5^{\circ}\times5^{\circ}$ at a resolution of $3'$, as was explained in Section \ref{sec:scale-invariant}. Therefore, since the first iteration of \forse{} allows one to go from maps with dimensions of $20^{\circ}\times20^{\circ}$ and a resolution of $80'$ to maps at a resolution of $12'$, the second iteration can produce  $5^{\circ}\times5^{\circ}$ maps at a resolution of $3'$. In order to preserve the proportions among all the relevant quantities (i.e., patch dimension, resolution of input, and resolution of output), the input patches for the second iteration should have dimensions of $5^{\circ}\times5^{\circ}$ and a resolution of $20'$. We can obtain those patches by smoothing and subdividing the output of \forse{}.

In practice, we pre-processed each of the 174 $M^{Q/U, 20^{\circ}}_{12\arcmin}$ maps with $320\times320$ pixels obtained from the first iteration in the following way:
\begin{enumerate}
\item We upsampled each map with $320\times320$ pixels to $1280\times1280$ by repeating each pixel four times. 
\item We smoothed these maps with a Gaussian kernel in order to reach a resolution of $20'$.
\item We subdivided each of these large squared patches into $5^{\circ}\times5^{\circ}$ patches, each one containing $320\times320$ pixels. We divided each large patch into 49 small ones, with a large overlap among them, made of 160 pixels on each side. This overlap is needed in order to avoid border effects when the composition and reprojection on the sphere is performed. 
\item At the end of this procedure, we have a set of $174\times49 = 8526$ patches for $Q$ and $U$, which is the total amount of patches covering the full sky and represents the $M^{Q/U, 5^{\circ}}_{20\arcmin}$ that will be used as the input to the second GAN iteration. 
\end{enumerate}

We applied the same pre-processing to the output of \forse{} and \forses{}12 in order to produce maps at $3'$ in both the deterministic case and the stochastic one. In the stochastic case, we added an additional random component, as is described in Section \ref{sec:stochasticity}. 

\subsubsection{Training and post-processing of \forsed{}} \label{sec:post}

Once the pre-processing steps described above are performed, the obtained 8526 patches can be used to train a new GAN model. However, the time cost is basically linear with the number of patches. We note that the 8526 patches have repeating pixels among them (see the steps above to get 8526 patches); thus, they are actually not independent. In order to save computational time, we fed as input to the generator, $G$, only a subset of 696 patches, randomly selected from the total 8526 patches. Once the network was trained, we applied it to the remaining patches.

\begin{figure*}[hbt]
    \centering
    \includegraphics[width=180mm]{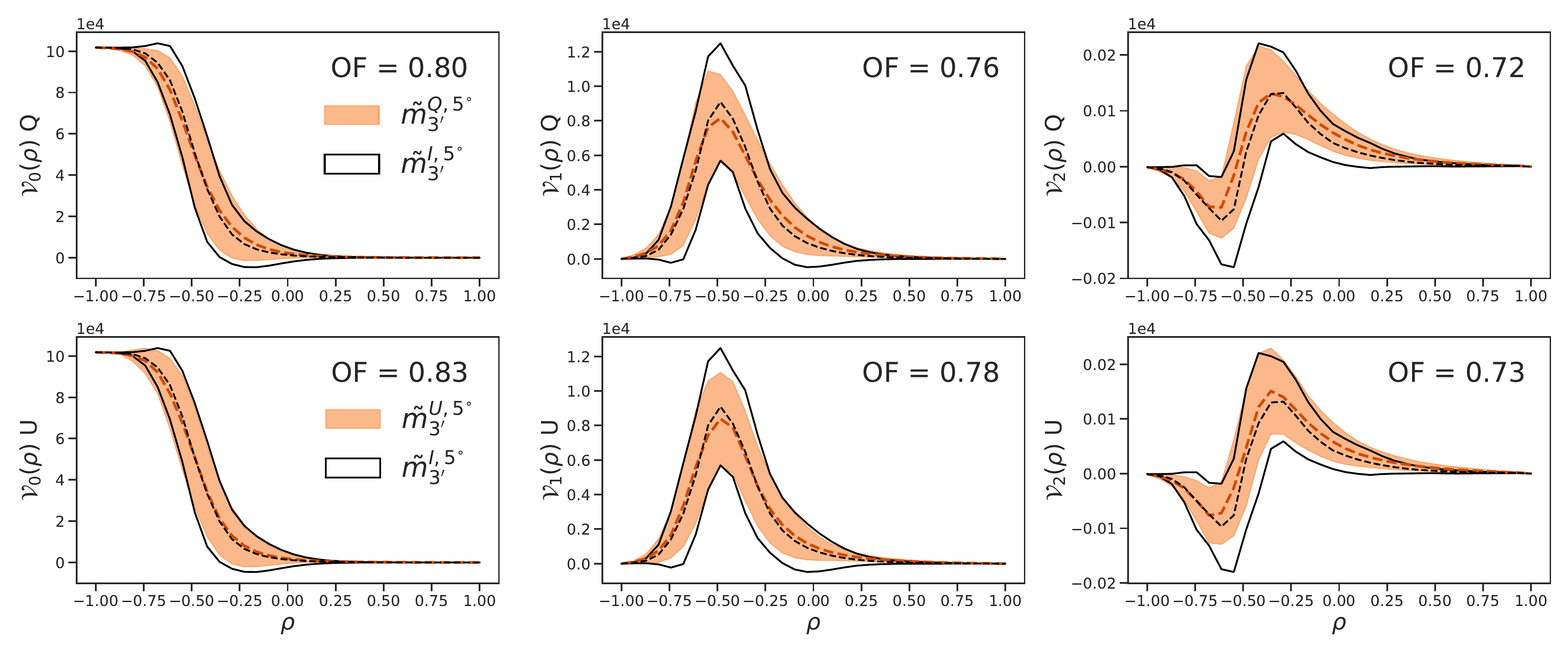}
    \caption{MFs overlapping between deterministic 3$\arcmin$ small scales and intensity small scales.}
    \label{fig:MF_deter}
\end{figure*}

In Figure \ref{fig:MF_deter} we show the MFs of the generated small scales at $3\arcmin$. The overlap with the target distribution is at a level of 70-80\% in the deterministic case. In the stochastic case (not shown), on the other hand, the overlap ranges between 60\% and 70\%, showing therefore that we are also able to generate non-Gaussian small-scale features at this higher resolution.

The output of the GAN models are patches, $\tilde{m}^{Q/U, 5^{\circ}}_{3\arcmin}$, containing the small-scale features that, as in the first iteration, have pixel values in the range of $[-1, 1]$. We therefore normalized them in physical units before multiplying them with the large-scale maps at the resolution of $20'$, obtaining 8526 $M^{Q/U, 5^{\circ}}_{3\arcmin}$ maps. 

\begin{figure}[hbt]
    \centering
    \includegraphics[width=0.35\textwidth]{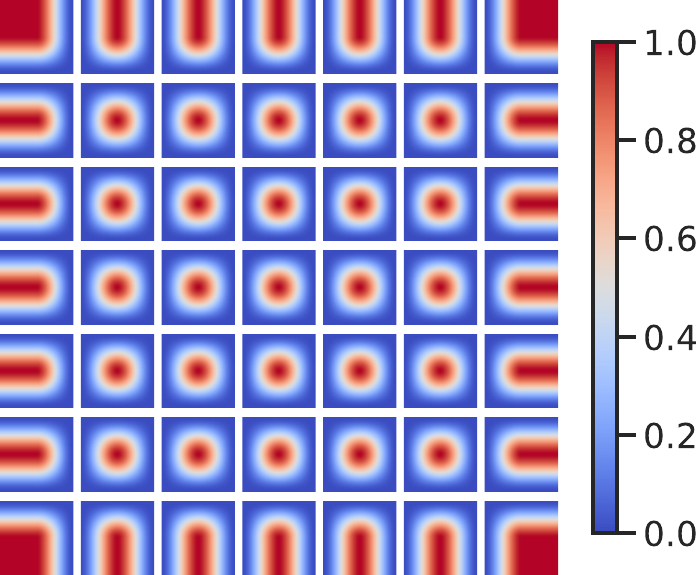}
    \caption{Apodized window function of each sub-patch at different positions to compose 49 sub-patches with a side length of $5^\circ$ into a patch with a length of $20^\circ$.} 
    \label{fig:composing}
\end{figure}

We recall that each of these patches comes from a subdivision of the larger maps with physical dimensions of $20^{\circ}\times20^{\circ}$ (as is described in Section \ref{sec:pre-deter}) into 49 sub-patches with a large overlap among each other. Therefore, before reprojecting them on the sphere to obtain the full-sky maps, we needed to recombine them. In order to avoid border effects, we did this by using $\cos^2$ apodization window functions, as is shown in Figure \ref{fig:composing}: each sub-patch was weighted with the corresponding window function, then they were summed together to form the 174 $M^{Q/U, 20^{\circ}}_{3\arcmin}$ maps that could then be reprojected on the sphere.

\subsubsection{Results of \forsed{} and \fsth{} at $3\arcmin$}

\begin{figure*}[hbt]
    \centering
    \includegraphics[width=0.85\textwidth]{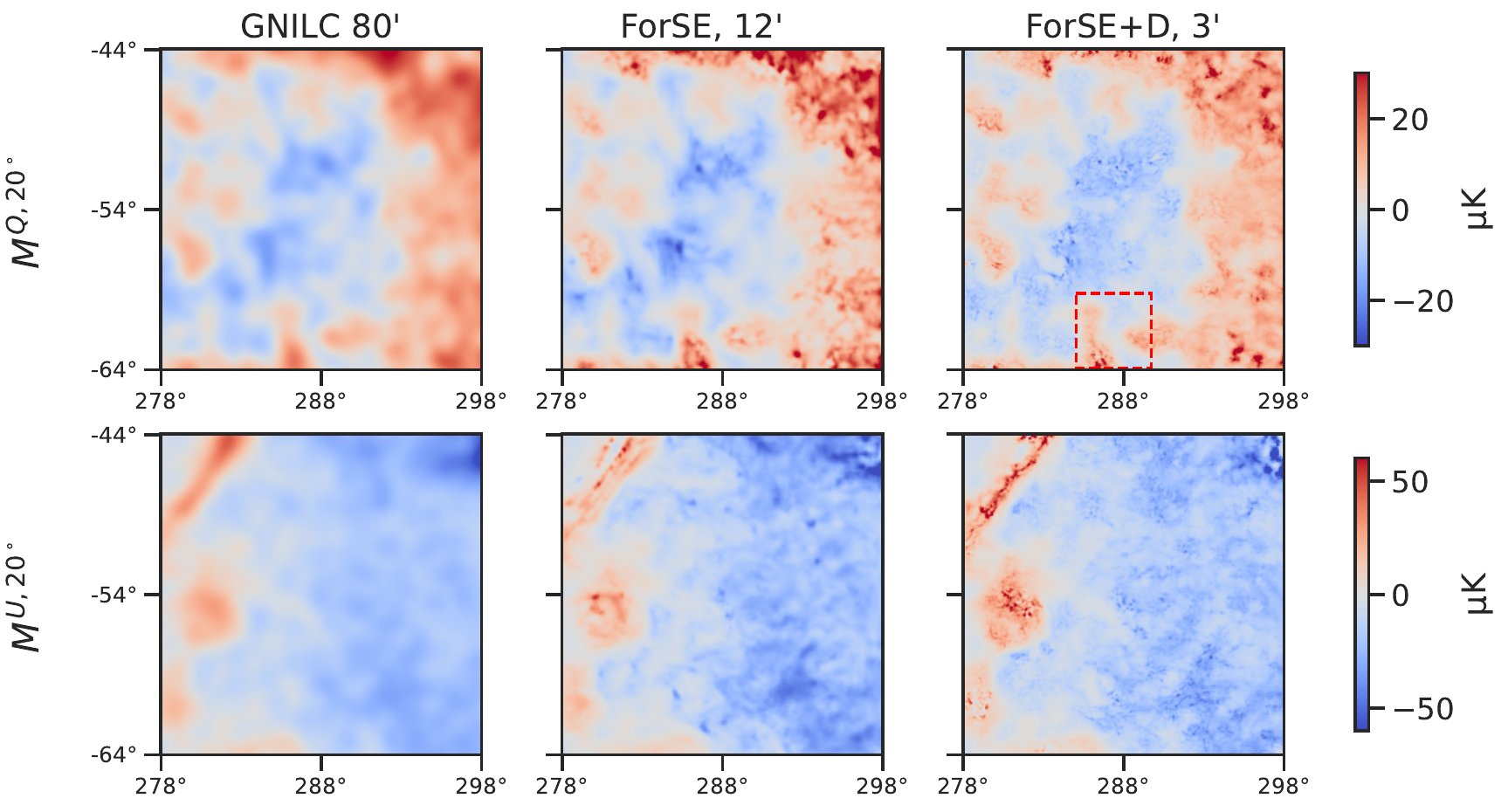}
    \caption{$20^\circ \times 20^\circ$ patches of three different resolutions. From left to right are GNILC maps at $80\arcmin$, \forse{} maps at $12\arcmin$, and \forsed{} maps at $3\arcmin$.}
    \label{fig:patch_20_3amin}
\end{figure*}

\begin{figure*}[hbt]
    \centering
    \includegraphics[width=0.9\textwidth]{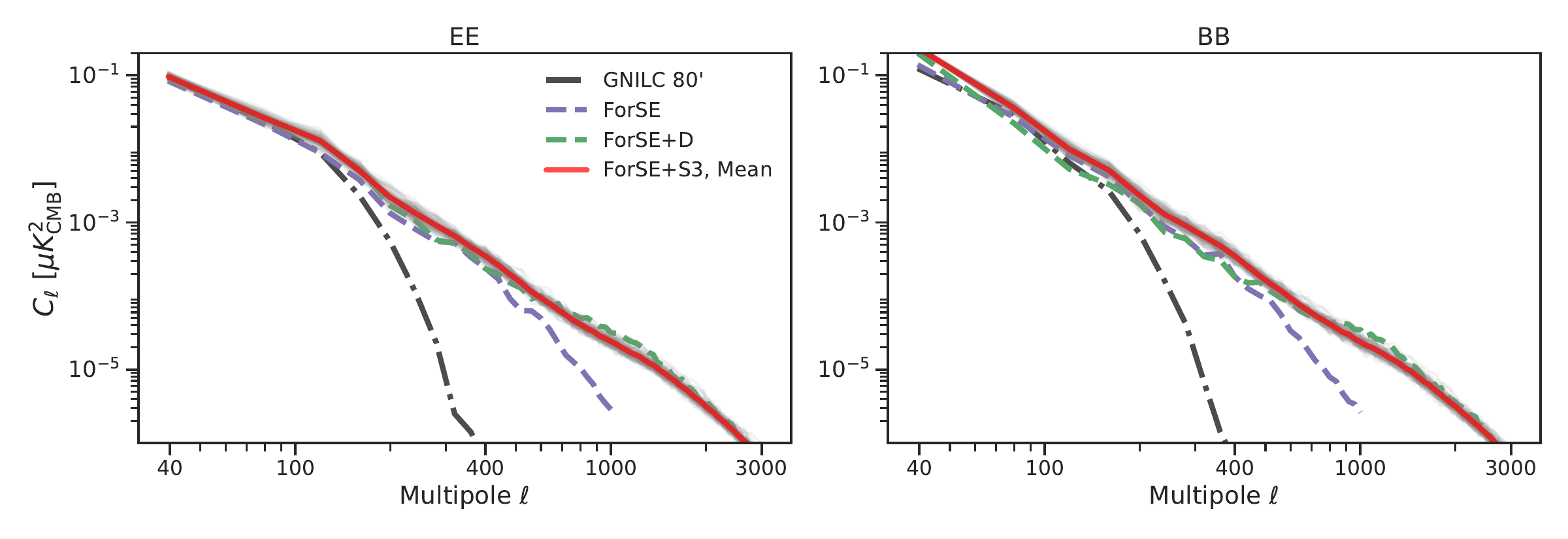}
    \caption{EE and BB power spectra of the squared patches at $3\arcmin$ shown in Figure \ref{fig:patch_20_3amin}. Dash-dotted black lines are the power spectra for GNILC $80\arcmin$ patches and purple lines are for the deterministic $12\arcmin$ patches. Green lines depict the behavior of \forsed{} maps, while gray lines show the power spectra of all the 100 realizations from \fsth{} and red lines are the means. }
    \label{fig:ps_3}
\end{figure*}

\begin{figure}[hbt]
    \centering
    \includegraphics[width=0.45\textwidth]{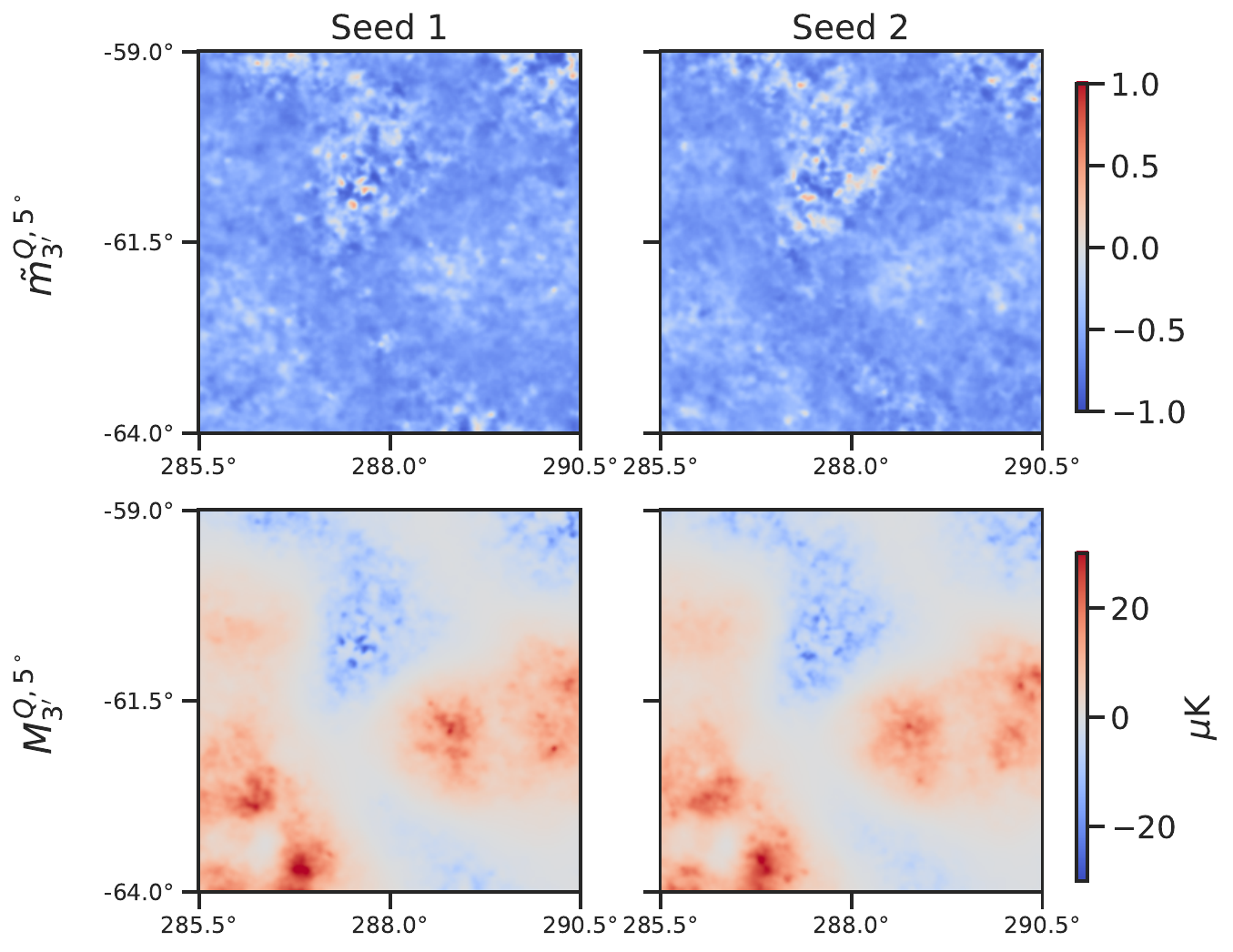}
    \caption{Upper panel: Two stochastic realizations of $Q$ maps, $\tilde{m}_{3\arcmin}^{Q, 5^\circ}$, from \fsth{}. Lower panel: $M_{3\arcmin}^{Q, 5^\circ}$ maps of the upper panel after the normalization step. This patch is at the position of the dashed red box in the upper right plot of Figure \ref{fig:patch_20_3amin}.}

    \label{fig:maps_5by5}
\end{figure}

In Figure \ref{fig:patch_20_3amin}, we show the comparisons of a selected patch at $80\arcmin$, $12\arcmin$ from \forse{} and maps at $3\arcmin$ from \forsed{}, from left to right. It is clear that \forsed{} can inject small-scale structures, following the modulation of the large-scale emission.

The power spectra, computed on the same patch, are shown in  Figure \ref{fig:ps_3}. As can be seen, the amplitude at low $\ell$ matches the one from the low-resolution GNILC maps, and the generated small scales have power at higher multipoles, with no breaks in the power spectrum that follows a power law, as was expected. 

\fsth{} was utilized to generate stochastic small scales at $3 \arcmin$, as is mentioned in \ref{sec:stochasticity}. Two realizations of small scales for the $Q$ map in the range of [-1, 1] with a side length of $5^\circ$ and centered on $[288^\circ, -61.5^\circ]$ (i.e., at the position of the dashed red box in the upper right plot of Figure \ref{fig:patch_20_3amin}) are shown in the upper panel of Figure \ref{fig:maps_5by5}, and in the lower panel we show the normalized maps combined with the large scales. The differences between the two realizations should be noted and both of them trace the large-scale features of the input maps.  We also show the power spectra of 100 realizations in Figure \ref{fig:ps_3}, where we also plot the mean, validating the results of stochasticity on the power spectrum level. 

\section{Validations of full-sky maps from \forsep{}} \label{sec:full-sky}

In this section, we show the maps and power spectra after reprojecting the flat patches back to the sphere, following the algorithm in the appendix of \citetalias{2021ApJ...911...42K}. Before showing the results, we introduce a further step to adjust the E-to-B ratio in the simulated maps to match the observations at large scales. 

\subsection{Fine-tuning of the E-to-B ratio to match observations}

 \begin{figure*}[hbt]
    \centering
    \includegraphics[width=1.0\textwidth]{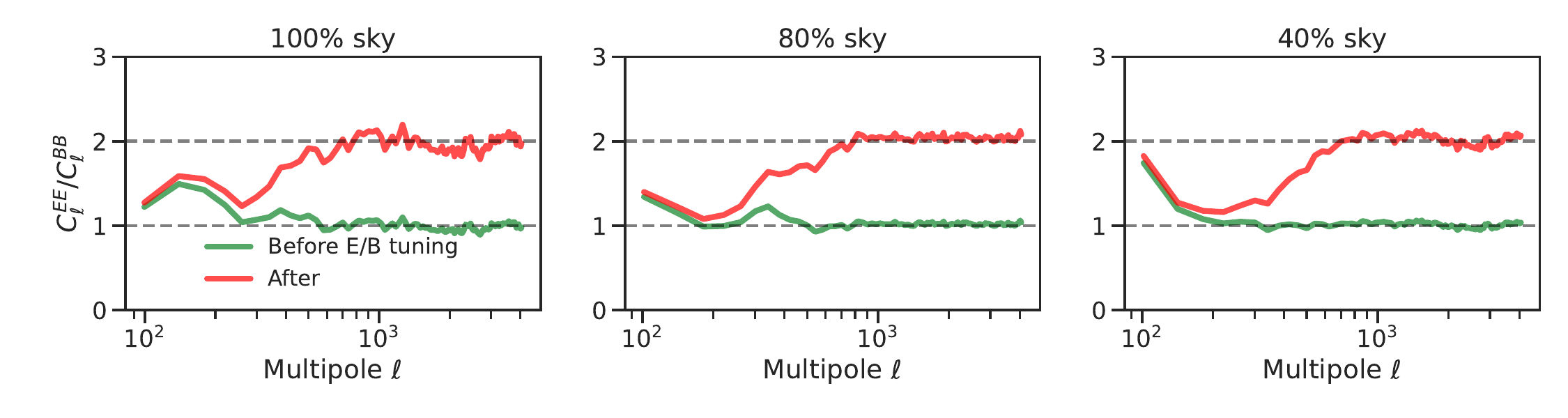}
    \caption{ E-to-B ratios of non-tuned and tuned maps in green and red lines, for different sky masks.}
    \label{fig:e2b}
\end{figure*}

We recall that, due to the limitation of observational data in polarization, we used the statistical properties of the total intensity small scales to be the ground truth of both $Q$ and $U$  during the training process. This implies that the output high-resolution map will have the same power for $E$ and $B$ modes. However, high-frequency observations of the Planck satellite have shown the existence of an asymmetry between the measured power of thermal dust emission in the $E$ and $B$ modes, with $A_{BB}/A_{EE} \sim 0.5$ over the multipole range of $40 \le \ell \le 600$ \citep{2020A&A...641A..11P}. Therefore, in order to match the real observations, we applied a fine-tuning to the full-sky maps, obtained after all the steps mentioned above. 

We first transformed $QU$ maps into the harmonic space to obtain the $a_{\ell m}^{E}$ and $a_{\ell m}^{B}$ coefficients. Then we applied a factor of $\sqrt{0.5}$ to the $a_{\ell m}^{B}$ to decrease its power, since $a_{\ell m}^{E}$ and $a_{\ell m}^{B}$ have the same variance as was just mentioned. Finally, we transformed the tuned $a_{\ell m}s$ back to the map space to get maps that match the observed ratio of 0.5 at the power spectra level. We note that the large scales in the output maps of \forse{}/\forsep{} remain to be the observed ones, so the tuning was applied only to the small scales; that is, $a_{\ell m}s$ belonging to $\ell > 135$\footnote{Corresponding to 80$\arcmin$: $\ell = \pi / \theta $ = $\pi / (80/60/180*\pi) = 135$} and the transition between large and small scales was smoothed with a sigmoid function to avoid discontinuities in the final power spectra.

The E-to-B ratios of the power spectra calculated for different fractions of sky are shown in Figure \ref{fig:e2b}. The red lines show the ratios after the tuning steps, which were indeed corrected to $\sim 2$ for the injected small scales.

\subsection{Full-sky maps}

\begin{figure*}[hbt]
    \centering
    \includegraphics[width=120mm]{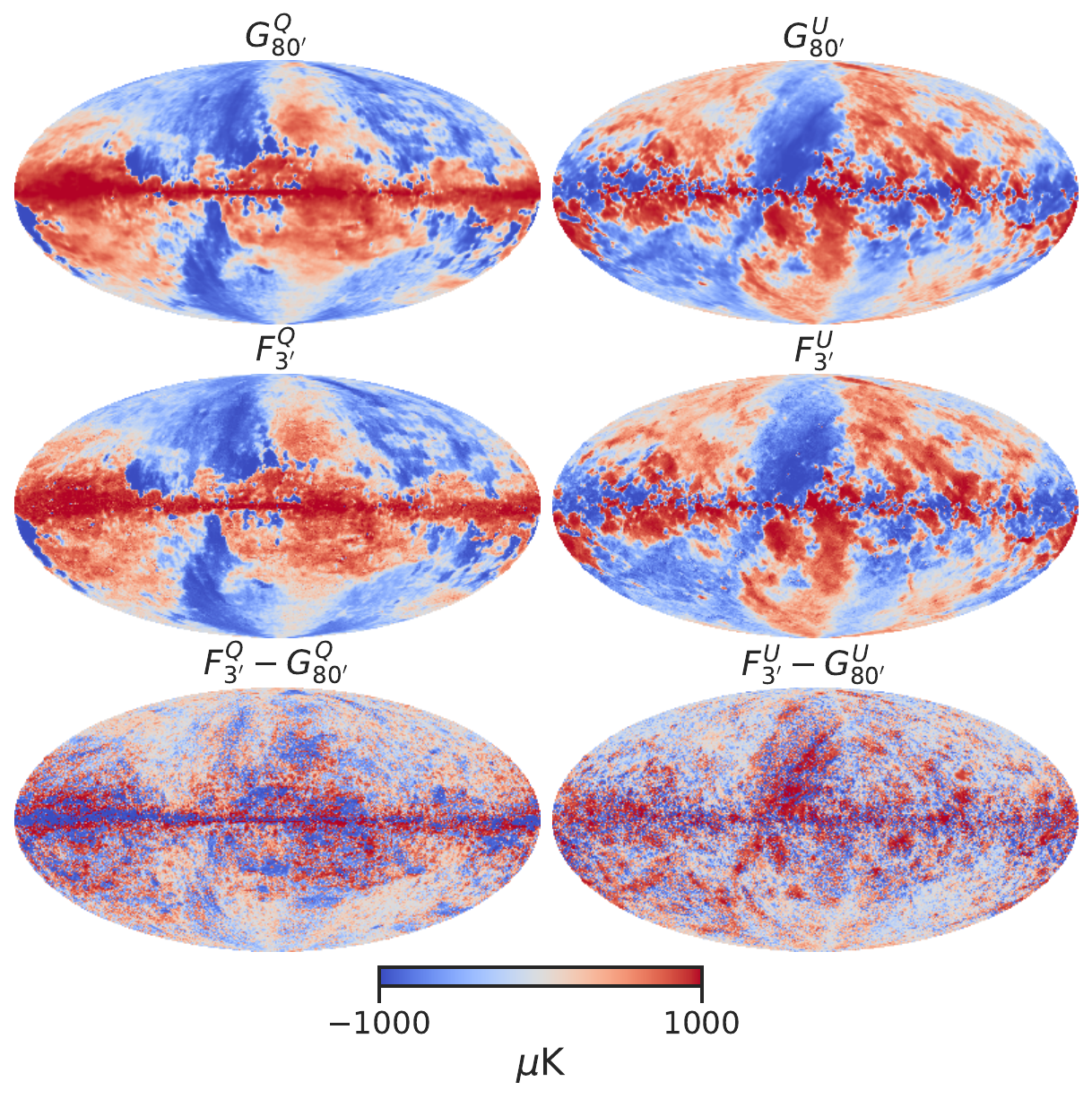}
    \caption{Top panel: Full-sky polarization maps (left: Stokes $Q$, right: Stokes $U$) for the GNILC template at a $80\arcmin$ angular resolution. These maps are the input to our algorithm. Middle panel: Maps with small-scale features, up to $3\arcmin$, generated by \forsed{}. Bottom panel: The difference between the two maps. The residuals mostly encode smaller angular scales, as was expected.}
    \label{fig:full_sky}
\end{figure*}

We are now ready to present the full-sky maps and specifically, we show the \forsed{} maps at $3\arcmin$, at Nside = 4096, in Figure \ref{fig:full_sky}, compared with the input GNILC ones at $80\arcmin$. The difference between the two highlights the presence of small-scale features in the \forsed{} maps. When zooming into a patch that consists of two $M^{Q/U, 20^{\circ}}_{3\arcmin}$ patches we also find no clear edge effects. These results show that border effects from reprojection are negligible.

\begin{figure*}[hbt]
    \centering
    \includegraphics[width=160mm]{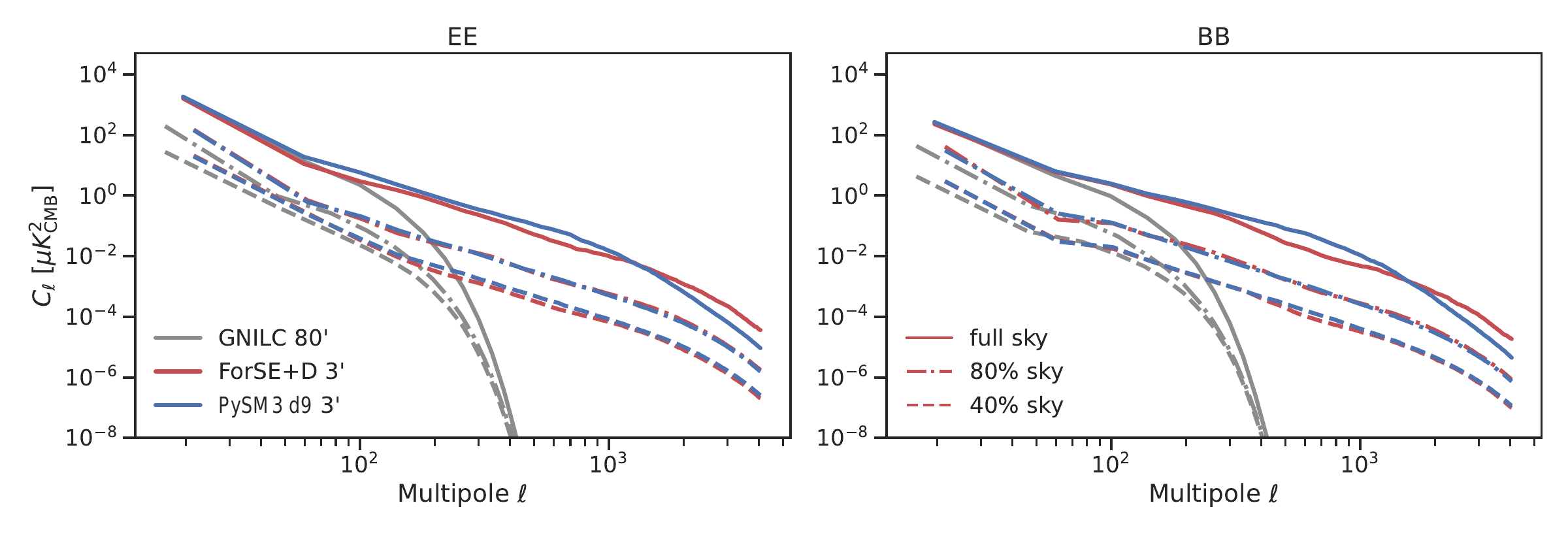}
    \caption{EE and BB Power spectra of GNILC $80\arcmin$, deterministic $3\arcmin$ in gray and red, with $f_{sky}$ = 1, 0.8, and 0.4 in solid, dash-dotted, and dashed lines, respectively. In order to make a comparison, we also show the power spectra of $\texttt{PySM3}$ \texttt{d9} dust map at 353GHz in blue.}
    \label{fig:cl_forse}
\end{figure*}

In Figure \ref{fig:cl_forse}, the power spectra of the maps on the sphere at different resolutions are shown for different sky fractions. As was expected, our maps at $3\arcmin$ can further extrapolate the power up to $\ell \sim 3600$, which corresponds to the drop scale of $3\arcmin$. We also note the absence of discontinuities at the transition from large scales to small scales ($\ell>200$), implying that the small scales are produced with a similar pattern as the one at large scales, attributed to the normalization step to rescale the generated small scales in the range of [-1, 1] to the correct amplitude, as was mentioned in Section \ref{sec:post}. In order to make a comparison with the latest \texttt{d9} dust model from $\texttt{PySM3}$\footnote{\url{https://github.com/galsci/pysm}} \citep{panex}, we also show the power spectra of \texttt{d9} maps at 353GHz in blue. Although the latter was produced with entirely different methods, the power spectra of \forsed{} maps at $3\arcmin$ are impressively close to those of \texttt{d9} maps, even for different sky fractions.

\begin{figure}[hbt]
    \centering
    % \vspace{0.15cm}
    \includegraphics[width=85mm]{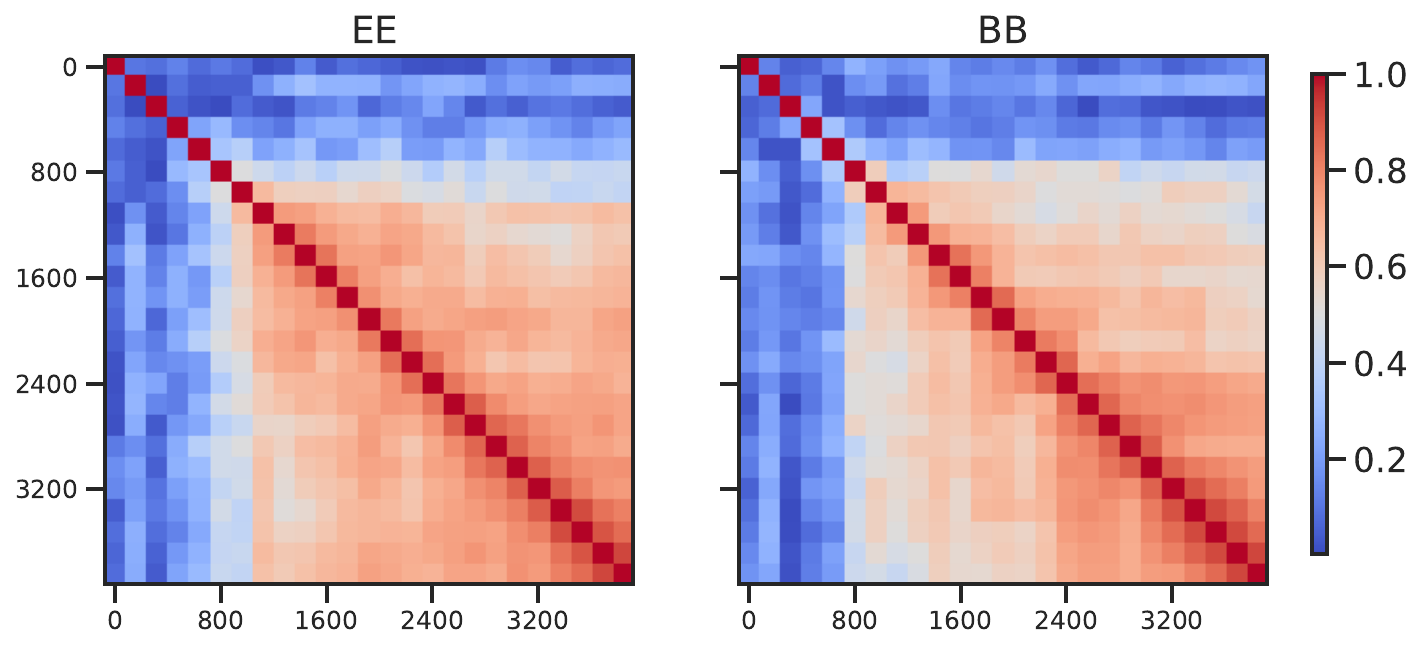}
    \vspace{0.05pt}
    \caption{Covariance matrices of maps at $3\arcmin$ on the sphere with an $80\%$ sky mask, shown in absolute values. Calculated for power spectra of an $\ell$-bin range of [40, 4096] with a bin width of 160. Non-diagonal terms are normalized with the diagonal values, i.e., $ R_{ij} = \frac{ C_{ij} } { \sqrt{ C_{ii} C_{jj} } }$.}
    \label{fig:cov_3_full}
\end{figure}

We also calculated the power spectra of 100 realizations of full-sky maps at $3\arcmin$, generated with \fsth{}, with a $80\%$ sky mask, up to $\ell_{max} = 4096$ and with a bin width of $\Delta \ell = 160$. The covariance matrix from these 100 realizations is shown in Figure \ref{fig:cov_3_full}. The correlation among multipoles at small scales is further evidence that the small scales at $\ell>800$ were synthesized in a non-Gaussian way. In fact, if the small-scale features were produced with Gaussian properties, we would not observe any non-diagonal correlation, which is verified from our experiment to calculate the covariance matrix of a purely Gaussian field. We devote the next section to further deepening the non-Gaussian properties of our maps.

\subsection{Non-Gaussianity measured on the sphere} \label{sec:non-gaussianity}

In previous sections, we considered the MFs in order to characterize non-Gaussianity for flat patches. Here, we expand the analysis to the full-sky maps, by exploiting the algorithms described in \cite{2022OJAp....5E..13G}, in order to calculate MFs for spherical maps in the \texttt{HEALPix} format. We focus on the small scales injected by \forse{} and \forsed{} and here we used the multipole range [200, 2048] to band-pass filter the raw maps and applied a $80\%$ sky mask to ignore the Galactic plane. 

\begin{figure*}[hbt]
    \centering
    \includegraphics[width=180mm]{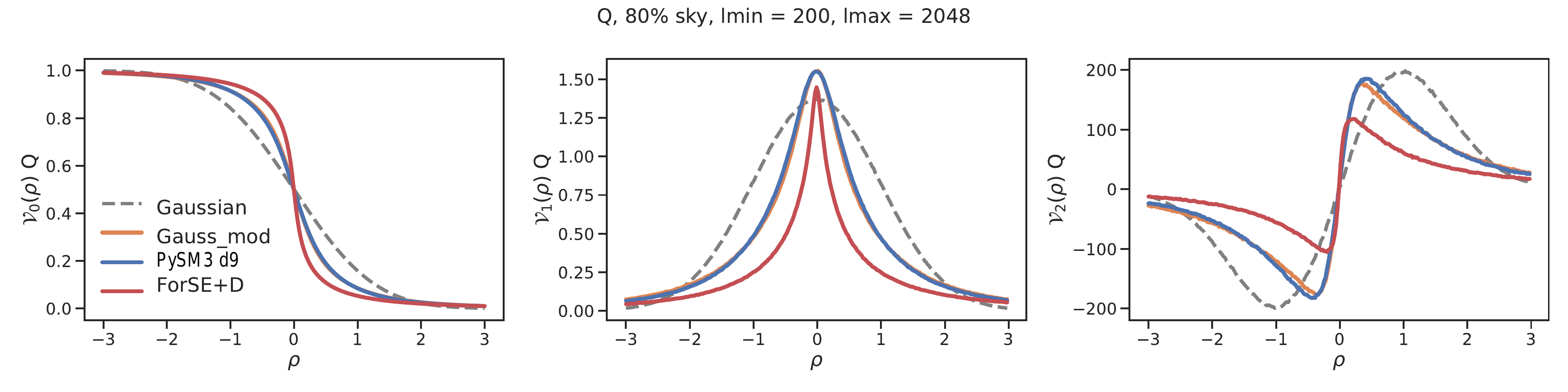}
    \caption{The three kinds of MFs for the \forsed{} $Q$ full-sky map at $3\arcmin$ with an $80\%$ sky mask (red). The raw map is band-pass filtered to keep small scales only, corresponding to the multipole range [200, 2048]. The dashed gray lines show the MFs of the Gaussian maps. Results for $\texttt{PySM3}$ \texttt{d9} and modulated Gaussian maps are also shown in blue and orange, respectively.}
    \label{fig:MFs_full}
\end{figure*}

In Figure \ref{fig:MFs_full} we show the results of the MFs statistics applied to the sphere, confirming our previous claims. Differences in the \forsed{} map are visible with respect to the case of Gaussian maps in dashed gray for all three kinds of MFs. The Gaussian maps were generated from a random realization from the  power spectra of the \forsed{} $Q$ map. We also show the results of the latest $\texttt{PySM3}$ \texttt{d9} dust maps in blue, which exhibits a deviation from Gaussianity, although being similar to the modulated Gaussian maps in orange, whose non-Gaussianity is supposed to derive only from the modulation of large scales. We further checked that the results are robust for the $40\%$ sky mask and for $\ell_{min}$ = 500 and 1000. 

\section{Discussion} \label{sec:conclusions}

In this paper, we extend the ability of the \forse{} model proposed in \citetalias{2021ApJ...911...42K} based on the GAN technique to simulate non-Gaussian stochastic small scales of polarized thermal dust emission up to $3\arcmin$. We have trained three new models, which are summarized in Table \ref{tab:summary}, together with the one trained in \citetalias{2021ApJ...911...42K}. 

Based on the results obtained in \citetalias{2021ApJ...911...42K}, our first test was to bring stochasticity into our models by adding a random component into the input and to train a new network called \fstw{} so that it can generate different maps with different seeds. Minkowski functionals were used to quantify the level of non-Gaussianity in the maps and in Figure \ref{fig:MFs_random_12} we demonstrate that the polarized thermal dust small scales have a similar level of non-Gaussianity to that in the intensity small scales. Different realizations of maps at $12\arcmin$ for a specific patch and the corresponding power spectra are shown in Figure \ref{fig:ps_12}, which indeed have expected variations at small scales and the correct amplitude scaling with multipoles for power spectra. 

We then considered the case of a $3\arcmin$ angular scale, where lensing $B$-modes have the strongest signal, and we studied the deterministic case first, \forsed{}. By relying on the scale invariance assumption discussed in Section \ref{sec:scale-invariant} and the pre- and post-processing steps outlined in Figure \ref{fig:forse_flowchart}, we trained the model to generate maps at $3\arcmin$ out of those at $12\arcmin$, which are the output from \forse{}. MFs distributions in Figure \ref{fig:MF_deter} represent a verification of the training process. Maps and power spectra of a selected patch of maps at $3\arcmin$ are shown in Figure \ref{fig:patch_20_3amin}, where the $3\arcmin$ power is present as expected. \fsth{} was finally trained in order to generate stochastic small scales at $3\arcmin$. Two realizations of small scales are shown in Figure \ref{fig:maps_5by5}, with their power spectra shown in the right panel of Figure \ref{fig:patch_20_3amin}, showing the expected multipole scaling.

After obtaining the small scales on flat patches, we reprojected them onto the celestial sphere in order to get full-sky maps in the \texttt{HEALPix} format, with $N_{side} = 2048$ for maps at $12\arcmin$ and 4096 for maps at $3\arcmin$. We show the Mollview projection of the deterministic maps at $3\arcmin$ in Figure \ref{fig:full_sky} and compare them with the observed \planck{} GNILC maps, validating both the effectiveness of the injected small scales and the reprojection process. The power spectra of deterministic maps at $80\arcmin, 12\arcmin$, and $3\arcmin$ with $100\%, 80\%$, and $40\%$ sky masks are plotted in Figure \ref{fig:cl_forse}, indicating that the small scales generated preserve the same anisotropies as the low-resolution observations. 

We further obtain the covariance matrix of power spectra up to $\ell_{max} \sim 4000$ from 100 realizations of maps at $3\arcmin$ in Figure \ref{fig:cov_3_full}, which has a strong correlation between different multipoles at small scales, and thus highlights the non-Gaussianity in our maps. We note that for a Gaussian field, the variances along the diagonal line are smaller than the variances we obtained here and the off-diagonal correlation is zero. We repeated the calculation of the covariance matrix for the \texttt{PySM3 d11} model, which is also designed to generate multiple realizations of small scales of thermal dust emission \citep{panex}, and find that off-diagonal elements are also close to zero, meaning that the level of non-Gaussianity of the injected small scales on the full sky is small. 

Finally, we measured the level of non-Gaussianity using MFs in the simulated full-sky maps in Figure \ref{fig:MFs_full}, and compared it with the small scales simulated within the Gaussian assumption. The difference between the MFs obtained from the two kinds of maps is another clear indication of the non-Gaussian small-scale component generated with \forsep{}.

Once the steps above are achieved, we can use the synthetic dust Stokes $Q$ and $U$ maps at 353GHz at $3 \arcmin$ resolution as a template and appropriately scale across different frequencies by taking the observed dust SEDs. The series of maps can then serve as a simulation suite to obtain the scatter of component separation against variations in the foreground realization. Moreover, high-resolution maps are essential in order to validate the lensing reconstruction methods, tested so far on foreground models where the arcminute structure is either absent or Gaussian \citep{2023arXiv231205184L}, or based on models of the Galactic magnetic fields in order to reproduce a non-Gaussian pattern \citep{2020JCAP...06..030B}. In a forthcoming work, we will investigate if any variation of the results shown in \cite{2020JCAP...06..030B} occurs when considering the foreground maps exhibiting non-Gaussianity generated through GANs in this paper. 

With the new version of $\texttt{PySM3}$ under development, we capitalize on integrating our maps into a new model of the latest $\texttt{PySM}$ packages for public use. Maps will be shared upon request to the authors. The code to generate maps has also been made publicly available.\footnote{\url{https://github.com/yaojian95/ForSEplus}.} 
Finally, we report here future improvements in the \forse{}  algorithm. First, more observations are needed. In fact, the assumptions made in Section \ref{sec:ForSEnew} are somewhat a compromise due to the lack of observation at the required resolution. The models will get more reliable as more observations become available. Second, by considering the network architecture, the loss function accounting for the non-Gaussianity of the targets produced by the GAN may be considered, as in the current implementation this process is not automatized. A way to quantify the level of non-Gaussianity with a formalism that is differentiable with respect to the input pixels would be desirable, as then we could construct the loss function of the network in order to include the information of non-Gaussianity, which will effectively guide the generator. The last point we want to mention is that the operation of adding noise is to some extent like the training process of diffusion models in deep learning, which is more natural than what is done in this work, so if we turn to use the network of diffusion model, it may have a better performance.

\section*{Acknowledgement}
The authors acknowledge partial support by the Italian Space Agency LiteBIRD Project (ASI Grants No. 2020-9-HH.0 and 2016-24-H.1-2018), as well as the InDark and LiteBIRD Initiative of the National Institute for Nuclear Phyiscs, and the RadioForegroundsPlus Project (HORIZON-CL4-2023-SPACE-01, GA 101135036). G.P. acknowledges support from Italian Research Center on High Performance Computing Big Data and Quantum Computing (ICSC), project funded by European Union - NextGenerationEU - and National Recovery and Resilience Plan (NRRP) - Mission 4 Component 2 within the activities of Spoke 3 (Astrophysics and Cosmos Observations). J.Y. thanks Yun Zheng for useful discussions. This research used resources of the National Energy Research Scientific Computing Center (NERSC), a U.S. Department of Energy Office of Science User Facility located at Lawrence Berkeley National Laboratory, operated under Contract No. DE-AC02-05CH11231.

\textit{Software used}: Astropy \citep{2013A&A...558A..33A,2018AJ....156..123A}, PySM \citep{Zonca_2021, 2017MNRAS.469.2821T}, NumPy \citep{harris2020array}, Namaster \citep{2019MNRAS.484.4127A}, numba \citep{lam2015numba}, Tensorflow \citep{tensorflow2015-whitepaper}, reproject \citep{thomas_robitaille_2023_7584411}, HEALPix \citep{Zonca2019, 2005ApJ...622..759G}

\bibliographystyle{aa} 
\bibliography{references}

\end{document}